\documentclass[preprint,showpacs,preprintnumbers,amsmath,amssymb]{revtex4}
\usepackage{graphicx}% Include figure files

\begin{document}

\title{Proton transport and torque generation in rotary biomotors}
\author{ A. Yu. Smirnov$^{1}$, S. Savel'ev$^{1,2}$,  L. G. Mourokh$^{1,3}$, and Franco Nori$^{1,4}$ }

\affiliation{ $^1$ Advanced Science  Institute, The Institute of Physical and
Chemical Research (RIKEN), \\
Wako-shi, Saitama, 351-0198, Japan \\
$^2$ Department of Physics, Loughborough University, Loughborough
LE11 3TU, UK \\
$^3$ Department of Physics, Queens College, The City University of
New York, Flushing, New York 11367, USA \\
$^4$ Center for Theoretical Physics, Physics Department, The
University of Michigan, Ann Arbor, MI 48109-1040, USA}

\date{\today}

\begin{abstract}
{We analyze the dynamics of rotary biomotors within a simple
nano-electromechanical model, consisting of a stator part and a
ring-shaped rotor having twelve proton-binding sites. This model is
closely related to the membrane-embedded F$_0$ motor of adenosine
triphosphate (ATP) synthase, which converts the energy of the
transmembrane electrochemical gradient of protons into mechanical
motion of the rotor.  It is shown that the Coulomb coupling between
the negative charge of the empty rotor site and the positive stator
charge, located near the periplasmic proton-conducting channel
(proton source), plays a dominant role in the torque-generating
process. When approaching the source outlet, the rotor site has a
proton energy level higher than the energy level of the site,
located near the cytoplasmic channel (proton drain). In the first
stage of this torque-generating process, the energy of the
electrochemical potential is converted into potential energy of the
proton-binding sites on the rotor. Afterwards, the tangential
component of the Coulomb force produces a mechanical torque.  We
demonstrate that, at low temperatures, the loaded motor works in the
shuttling regime where the energy of the electrochemical potential
is consumed without producing any unidirectional rotation. The motor
switches to the torque-generating regime at high temperatures, when
the Brownian ratchet mechanism turns on. In the presence of a
significant external torque, created by ATP hydrolysis, the system
operates as a proton pump, which translocates protons against the
transmembrane potential gradient. Here we focus on the F$_0$ motor,
even though our analysis is applicable to the bacterial flagellar
motor.}
\end{abstract}

\pacs{87.16.A--,\ 87.16.Uv,\ 85.85.+j}

\maketitle

\section{Introduction}

Biological rotary motors, such as  the F$_0$ motor of ATP (adenosine
triphosphate) synthase  and the bacterial flagellar motor (BFM),
convert energy of the transmembrane electrochemical potential into
mechanical motion \cite{Alberts02,BrowneNat06}. The gradient of the
electrochemical potential in the living cells is maintained by the
metabolic mechanism that translocates ions (here protons) from the
negative (or cytoplasmic) side of the membrane to its positive
(periplasmic) side \cite{Alberts02,Pumps,Kim07,PRE08}. A mechanical
torque is generated when protons flow down the electrochemical
gradient across the mitochondrial or cell membranes
\cite{Glag78,Junge97,Elston98,Dimroth99,OsterWang03,Aksim04,Capaldi02,Dimroth06,Berg03,Xing06}.
Bacteria employs this torque directly for chemotaxis
\cite{Chemotaxis}, whereas in ATP synthase the rotational energy is
transmitted to the F$_1$ motor mechanically linked to the F$_0$
portion.  The out-of-membrane component F$_1$ contains three sites,
which harness the energy of mechanical rotations to catalyze the
synthesis of ATP molecules. In the following, we concentrate on the
F$_0$ motor, although our theoretical analysis can be applied to the
BFM as well, since these motors have many common features.

\subsection{Structure of the biomotor}

The F$_0$ motor is embedded into a plasma membrane of bacteria or
into an inner mitochondrial membrane. It consists of two components:
 (i) a stator or subunit $a$, tightly attached to the membrane, and (ii) a
cylinder-shaped rotor (subunit $c$), which can freely rotate around
its axis (Fig.~1). In the majority of models
\cite{Junge97,Elston98,Dimroth99,OsterWang03,Aksim04,Capaldi02,Dimroth06}
the rotor contains several (10 to 15) proton-binding sites located
in the middle of the membrane. The stator has two non-collinear
half-channels, which are perpendicular to the plane of Fig.~1. Each
one of these channels are connected either to the periplasmic side
of the membrane, having a higher electrochemical potential, or to
the cytoplasm, characterized by a lower potential. The source ($S$)
half-channel delivers protons from the periplasm (the positive, or,
P-side of the membrane) to the rotor-stator interface, and, finally,
to the proton-binding sites in the rotor plane. The drain ($D$)
channel translocates protons from this interface to the cytoplasmic,
or negative, N-side of the membrane. The rotor sites can be
protonated or deprotonated when they are in the immediate vicinity
of the source or drain channels. In ATP synthase from
\textit{Escherichia coli} the residues $c$Asp-61 are equally placed
along the rotating ring and serve as proton-binding sites carrying
protons from the source to the drain channel \cite{Aksim04}. An
empty rotor site has a negative charge.

\subsection{Motor mechanisms: Brownian ratchet and power stroke}

It is assumed \cite{Junge97,Elston98} that protons, delivered to the left side (Fig.~1) of the rotor-stator interface by the
source channel, bind and neutralize the empty rotor sites, and after that, travel clockwise (looking from the upper, periplasmic,
side of the membrane, as shown in Fig.~1a). Notice that all rotor sites outside of the rotor-stator interface are occupied with
protons and have no electric charge. The unprotonated negatively charged sites cannot face the lipid core of the membrane because
of a large ( $\geq$ 500 meV) desolvation penalty \cite{Armen06}. The proton escapes from the rotor during the process of
clockwise-directed rotor diffusion, when the corresponding site approaches the drain (cytoplasmic) channel located at the right
side of the rotor-stator strip (Fig.~1). The negatively-charged empty site is not able to move back now because of the strong
repulsion of the lipid medium. However, it can diffuse forward, in the clockwise (CW) direction, since there is no energy penalty
for the charge placed inside the hydrophilic rotor-stator interface. The site is populated and neutralized again when it reaches
the source channel at the left side of the interface, and the wheel can continue its CW rotation.

Because of random thermal forces, the rotor can also rotate in the counterclockwise (CCW) direction, and the populated site would
then approach the drain channel. In the most probable event, the site is depopulated and reflected from the lipid border of the
rotor-stator interface. However, with a smaller probability, the neutral site, occupied with a proton, crosses the border, and the
wheel rotates counterclockwise. When the population of the rotor sites occurs on the left side of the rotor-stator interface, and
the depopulation happens on the right side, the rotor diffusion is biased to rotate CW. This means that the F$_0$ motor works as a
Brownian ratchet \cite{Reimann02,Hanggi05,DMLRatch}.

References~\cite{Elston98,OsterWang03,WangOster02} have shown,
however, that the Brownian ratchet model is not able to explain the
operation of the F$_0$ motor in the presence of an external torque
(about 41 pN nm), which is produced by the F$_1$ motor and acts in
the opposite (CCW) direction. In addition to two half-channels, the
stator contains a positive charge (a residue $a$Arg-210 in
\textit{E.~coli}) which can be located between two proton-conducting
half-channels in the rotor plane \cite{Elston98,Aksim04}. As
schematically shown in Fig.~1a, upon passing the source channel, the
rotor site releases its proton and becomes negatively charged. This
charge is attracted by the positive residue, thereby generating a
torque moving the rotor in the CW direction (power stroke
component). Reference~\cite{Elston98} has demonstrated that the
positive stator charge is necessary for the efficient reverse
operation of ATP synthase, when it works as a proton pump under the
action of the external torque produced by ATP hydrolysis. The
irreplaceable role of the $a$Arg-210 residue for the ATP-driven
pumping activity of the F$_0$F$_1$ complex in \textit{E.~coli} has
been proven by direct mutagenesis studies \cite{Cain89}. It should
be noted that, despite the recent significant progress in
understanding the operation of ATP synthase, the physical mechanism
of the torque generation and proton translocation by the F$_0$ motor
calls for further investigation \cite{Capaldi02,Dimroth06}. Of
special interest is the possibility to mimic the main features of
this biological engine with the goal to create an artificial
nanomachine with an almost perfect energy conversion.

\subsection{Quantitative modelling}

Here we explore a simple model incorporating the most important aspects of biomolecular rotary motors
\cite{Junge97,Elston98,Dimroth99,OsterWang03,Aksim04} and employing a quantitative treatment based on methods of condensed matter
physics \cite{Wingr93,PRE08,RobPRB08}. These approaches have been previously applied to nano-electromechanical systems (NEMS) with
their mechanical motion affecting the electrical properties of electronic devices \cite{NEMS}. Similar processes take place in
nanoscale biological objects, where electrical and mechanical degrees of freedom are also strongly coupled, making them living
counterparts to artificial NEMS. Note that only nano-oscillators have been extensively studied by theorists, although a
single-molecule rotor and nanoelectromechanical rotational actuators \cite{GimZet} have been demonstrated experimentally. To the
best of our knowledge, no theoretical investigations of rotary NEMS have been reported yet. It can be expected that artificial
nanoengines built on the operating principles of biological motors could achieve the same level of efficiency and performance.

ATP synthase, along with many other enzymes \cite{Kim07}, should be considered as a nonequilibrium open system whose operation is
supported by a permanent flow of protons. To describe this flow we attach the F$_0$ motor to two proton reservoirs related to the
periplasmic (positive, P) side of the membrane and to the cytoplasmic (negative, N) side of the membrane. In the case of the
respiratory chain \cite{Alberts02}, the P-side and the N-side correspond to the intermembrane space and to the matrix,
respectively, divided by an inner mitochondrial membrane. Usually the proton transfer between the rotor sites and the P and N
aqueous sides of the membrane is represented by a set of Markov equations with a phenomenologically constructed transition matrix
\cite{Dimroth99,Xing06,WangOster02}. Here, with the methods of quantum transport theory \cite{Wingr93,NEMS,PRE08,RobPRB08}, we
derive the rate equations and provide simple expressions for transition coefficients, which explicitly depend on the difference
between electrochemical potentials of the proton reservoirs and on the distance between the proton-binding sites on the rotor and
on the stator. This approach gives a clear physical picture of the loading and unloading events as well as the torque-generation
process as a whole.

The paper is organized as follows. In Sec.~II we describe in detail
the model and formulate the Hamiltonian of the system. In Sec.~III
we present a Langevin equation for the mechanical motion of the
rotor and derive the rate equations for the populations of the
proton-binding sites on the rotor. In Sec.~IV we numerically solve
the equations of motion and discuss our results. Section V is
devoted to conclusions.

\section{Model}

We consider  twelve equally-spaced proton-binding sites ($\sigma
=1,\ldots,12$), attached to a ring-shaped rotor ($c$-subunit) of
radius $r_0$, at the points with the angular coordinates
$\phi_{\sigma}~=~\sigma \cdot \phi_0,$ where $\phi_0 = \pi/6$ is the
angular distance between the rotor sites. Then, if the rotor is
turned through an angle $\phi$, the proton-binding site $\sigma$ has
a coordinate $$\textbf{r}_{\sigma} = \{ r_0 \sin(\phi +
\phi_{\sigma}),\  r_0 \cos(\phi + \phi_{\sigma}) \}.$$ We choose the
coordinate origin at the center of the rotor wheel. As shown in
Fig.~1, the stator part of the motor (or $a$-subunit) contains the
source half-channel, $S$, which delivers protons from the P-side
(periplasm) of the membrane to the point with an angular position
$\phi_S = \pi$ on the rotor-stator interface. The drain
half-channel, $D$, also belonging to the stator, connects the point
$\phi_D = \pi - \phi_0 $ with the cytoplasmic (N) side of the
membrane. Also, the stator has a positive charge $q|e|$ located in
the rotor plane on the $y$-axis, with coordinates $\textbf{r}_q =
\{0,-r_0-d\}$.  Here $d$ is the distance from the charge $q|e|$ to
the closest proton-binding site on the rotor, $e$ is the electron
charge. As a result, the Hamiltonian of the system has the form:
\begin{eqnarray}
H = H_{\phi} + \sum_{k\alpha} \,E_{k\alpha}c_{k\alpha}^+c_{k\alpha} + H_{\rm tun} + \nonumber\\
\sum_{\sigma} E_{\sigma} n_{\sigma} + \sum_{\sigma} \,[ U
_{q}(\phi+\phi_{\sigma}) + U_{\rm con}(\phi+\phi_{\sigma}) ]\,(1-
n_{\sigma}), \label{H0}
\end{eqnarray}
where the Hamiltonian $H_{\phi}$ governs the mechanical motion of
the cylinder-shaped rotator, characterized by  the angle $\phi$,
counted clockwise from the $y$-axis. The occupation of the
proton-binding site $\sigma$, having eigenenergy $E_{\sigma}$, is
described by the creation and annihilation operators of protons,
$a_{\sigma}^+, a_{\sigma}$, with a corresponding population operator
$n_{\sigma} = a_{\sigma}^+a_{\sigma}.$ Fermi operators
$c_{k\alpha}^+,c_{k\alpha}$ are related to the state, with momentum
$k$, of the proton in the source and drain reservoirs $(\alpha =
S,D)$ with energy $E_{k\alpha} $. The Hamiltonian $H_{\rm tun}$
describes the proton transfer between the proton-binding sites on
the rotor and the source and drain channels on the stator.

\subsection{Coulomb interactions}

The empty site (with $n_{\sigma}=0$) has a negative charge, and,
because of this, it is attracted by the stator charge $q|e|$. The
screened Coulomb coupling between the negative charge, $-|e|$, of
the empty rotor site and the positive stator charge $q|e|$,
separated by the angular-dependent distance
$r_{q}(\phi+\phi_{\sigma}) = |\textbf{r}_{\sigma}-\textbf{r}_q|$, is
determined by the potential \cite{Aksim04}:
\begin{equation}
U _{q}(\phi +\phi_{\sigma})= -\ \frac{q e^2}{4\pi \varepsilon_0
\varepsilon r_{q}(\phi +\phi_{\sigma})} \exp\left(-\;
\frac{r_{q}(\phi +\phi_{\sigma})}{r_s}\right), \label{Uq}
\end{equation}
where $\varepsilon_0$ is the electric permittivity,  $\varepsilon$
is the dielectric constant of the medium, $r_s$ is the Debye
screening length, and
\begin{equation}
r_{q}(\phi+\phi_{\sigma}) =  \sqrt{ r_0^2 + (r_0+d)^2 + 2 r_0 (r_0 +
d) \cos(\phi+\phi_{\sigma})}. \label{rsigmaq}
\end{equation}

To take into account an energetic penalty for the charged rotor site
$\sigma$, when it faces a lipid bilayer outside of the rotor-stator
strip, we introduce a confinement potential,
\begin{equation}
U_{\rm con}(\phi+\phi_{\sigma}) = U_{c} \{ 1 - \exp[ -\lambda_c ( 1
- \cos(\phi + \phi_{\sigma} - \phi_c))^2 ] \}, \label{Usigmac}
\end{equation}
 where an angle $\phi_c$
corresponds to the middle of the rotor-stator strip, $\phi_c = \pi -
\phi_0/2,$ a parameter $\lambda_c$ is inversely proportional to the
width of the interface, and a potential $U_c$ represents a
desolvation penalty. The transfer of an ion with charge $-|e|$ from
water (with a dielectric constant $\varepsilon_1$) to the
hydrophobic membrane (having a constant $\varepsilon_2$) is subject
to the energy penalty \cite{Armen06},
\begin{equation}
U_c \ ({\rm meV}) = \frac{1440 \ e^2}{2 a
}\left(\frac{1}{\varepsilon_2} - \frac{1}{\varepsilon_1} \right),
\label{Uc}
\end{equation}
 where $a$ is the size of the cavity
(in nm) over which the charge is spread. For $\varepsilon_1=80,\
\varepsilon_2=3,$ and $a=0.2 $ nm, the desolvation energy is about
$U_c = 1160$ meV. We assume here that the proton-proton Coulomb
interaction between sites is small enough, so that the loading and
unloading of different sites occurs independently.

\subsection{Proton transfer}

 Protons in the source and drain
reservoirs are characterized by the Fermi distributions,
\begin{equation}
f_{\alpha}(\omega) =
\left[\exp\left(\frac{\omega-\mu_{\alpha}}{T}\right) +
1\right]^{-1}, \label{Fermi}
\end{equation}
 with temperature $T$ ($\hbar=1,\ k_B=1$) and electrochemical potentials $\mu_{S} = V/2, \mu_{D} = - V/2, $ where $V$ is the proton voltage build-up,
 in units of energy, meV. For ATP synthase
in \textit{Escherichia coli}, protons can be translocated from the
aqueous sides of the membrane to the rotor-stator interface by a set
of hydrogen-bonded chains \cite{Zundel95}. The mechanism of a proton
transfer between the rotor sites $c$Asp-61 and the terminal residues
of the source channel, $a$Asn-214, and the drain channel,
$a$Ser-206,  is not completely understood. However, molecular
dynamics simulations \cite{Aksim04} demonstrate that hydrogen bonds
can be formed between the proton-binding sites on the rotor
($c$Asp-61) and the terminal residues ($a$Asn-214 or $a$Ser-206).
These bonds are able to transfer protons within picoseconds either
by collective tunneling \cite{Zundel95}, or, which is more probable,
by classical hopping \cite{Elston98,Aksim04}. The proton
translocation process strongly depends on the distance between the
rotor sites and the stator residues and, at the molecular level, it
can be facilitated by internal rotations of transmembrane helices as
well as by the motion of other key elements \cite{Aksim04,
Dmitriev99,Rastorgi99}. In our simplified model, we mimic the
stator-rotor proton transitions with the Hamiltonian
\begin{equation}
H_{\rm tun} = - \sum_{k\alpha \sigma} [
\,T_{k\alpha}\,c_{k\alpha}^+\, a_{\sigma} +
T_{k\alpha}^*\,a_{\sigma}^+\, c_{k\alpha} ]\;w_{\alpha
}(\phi+\phi_{\sigma}), \label{Htun}
\end{equation}
which describes the effective proton tunneling with amplitudes
$T_{k\alpha}$, multiplied by factors $w_{\alpha
}(\phi+\phi_{\sigma})$. The factors $w_{\alpha
}(\phi+\phi_{\sigma})$ depend on the distance between the site
$\sigma$ and the final residue of the $\alpha$-channel located at
the point $\textbf{r}_{\alpha} = \{ r_0 \sin\phi_{\alpha},\, r_0
\cos\phi_{\alpha} \},$
\begin{equation}
|\textbf{r}_{\sigma}-\textbf{r}_{\alpha}| = \sqrt{2}r_0 \sqrt{ 1 -
\cos(\phi + \phi_{\sigma} - \phi_{\alpha})}.
\end{equation}
We approximate this dependence by the exponential function,
\begin{equation}
w_{\alpha }(\phi+\phi_{\sigma}) =
\exp(-\lambda_r|\textbf{r}_{\sigma}-\textbf{r}_{\alpha}|),
\end{equation}
characterized by a steepness $\lambda_r$. The value of $\lambda_r$
is inversely proportional to the size of the molecular groups
participating in the proton transport (either $c$Asp-61 and
$a$Asn-214, or $a$Ser-206). It should be noted that the specific
functional form of the factor $w_{\alpha \sigma}(\phi)$ is not
crucial for the model under study.

\section{Equations of motion}

The viscous medium creates a torque,
\begin{equation}
{\cal T}_r = \zeta_r \frac{d\phi}{dt}, \label{VisTorque}
\end{equation}
acting on the cylinder-shaped rotor, together with a stochastic
force $\xi$. For a cylinder with radius $r_0$ and height $h$,
rotating in a medium with viscosity $\eta$, the drag coefficient
$\zeta_r$ is defined by the formula \cite{Howard01,Nart06}:
\begin{equation}
\zeta_r = 4\pi \eta r_0^2 h. \label{zetaR}
\end{equation}
It follows from the Hamiltonian (\ref{H0}), that the processes of
loading and unloading of protons at the rotor-stator interface
generates an additional torque. Thus, the following Langevin
equation describes the biased overdamped diffusion of the rotor:
\begin{equation}
\zeta_r \dot{\phi} = \xi + {\cal T}_{\rm ext} - \sum_{\sigma} \,(1-
n_{\sigma})\,\frac{d}{d \phi}\, [ U _{q}(\phi+\phi_{\sigma}) +
U_{\rm con}(\phi+\phi_{\sigma}) ].
 \label{Langevin}
\end{equation}
Here $\xi(t)$ is a zero-mean value Gaussian fluctuation source,
characterized by the correlation function: $\langle \xi(t)
\xi(t')\rangle = 2 T \zeta_r \delta(t-t'),$ and ${\cal T}_{\rm ext}$
is an external torque produced by the F$_1$-motor. We expect that
the F$_0$ motor will rotate in the CW direction (looking from the
periplasm side of the membrane). Therefore, the torque from F$_1$,
which decelerates this motion,  should have a negative sign, ${\cal
T}_{\rm ext} < 0$. The effects of the rotor-stator proton
transitions on the mechanical motion, resulting from the Hamiltonian
$H_{\rm tun}$, are assumed to be negligibly small.

The  ``chemical" part of the problem, namely, the process of loading
and unloading the proton-binding sites on the rotor, is governed by
the Heisenberg equations for the population operators, $n_{\sigma} =
a_{\sigma}^+a_{\sigma},$ which can be derived \cite{PRE08,RobPRB08}
from Eqs. (\ref{H0}) and (\ref{Htun}):
\begin{equation}
\dot{n}_{\sigma} = i \sum_{k\alpha}\, [ T^*_{k\alpha}\, a_{\sigma}^+
c_{k\alpha} - T_{k\alpha} \,c_{k\alpha}^+ a_{\sigma} ] \,w_{\alpha }
(\phi+\phi_{\sigma}), \label{nsigma1}
\end{equation}
where the reservoir operators, $c_{k\alpha},$ are represented as a
sum of the free term, $c_{k\alpha}^{(0)},$ and the term describing
the reservoir response:
\begin{equation}
c_{k\alpha} = c_{k\alpha}^{(0)} - T_{k\alpha} \sum_{\sigma} \int
dt_1 \, g_{k\alpha}^r(t,t_1) a_{\sigma}(t_1) w_{\alpha \sigma}(t_1).
\label{ckalpha}
\end{equation}
Here $$g_{k\alpha}^r(t,t_1) = - i \exp[ - iE_{k\alpha}(t-t_1)]\,
\theta(t-t_1)$$ is the retarded Green function of the
$\alpha$-reservoir,  $\theta (t-t_1)$ is the Heaviside step
function, and $$w_{\alpha \sigma}(t)\equiv w_{\alpha
}(\phi(t)+\phi_{\sigma}).$$ The correlator of the free reservoir
operators is determined by the Fermi-distribution function
$f_{\alpha}(E_{k\alpha})$: $$\langle
c_{k\alpha}^{(0)+}(t)c_{k\alpha}^{(0)}(t_1)\rangle =
f_{\alpha}(E_{k\alpha})\ \exp[ iE_{k\alpha}(t-t_1)].$$ On
substituting Eq.~(\ref{ckalpha}) into Eq.~(\ref{nsigma1}), and
averaging the latter equation over the fluctuations of the proton
reservoirs, we obtain
\begin{eqnarray}
\langle \dot{n}_{\sigma}\rangle + \sum \Gamma_{\alpha} w_{\alpha \sigma}^2(t) \,\langle n_{\sigma}\rangle  = \nonumber\\
\sum_{\alpha} |T_{k\alpha}|^2\int dt_1 \,f_{\alpha}(E_{k\alpha})\,
e^{iE_{k\alpha}(t-t_1)} \,w_{\alpha \sigma}(t)\,w_{\alpha
\sigma}(t_1) \times  \nonumber\\
\langle [a_{\sigma}(t), a_{\sigma}^+(t_1)]_+ \rangle \,\theta(t-t_1)
+ h.c. \label{nsigma2}
\end{eqnarray}
We introduce here the transition rates between the proton-binding
sites and the reservoirs, $$\Gamma_{\alpha} = 2\pi \sum_k
|T_{k\alpha}|^2 \delta (E - E_{k\alpha}),$$ which are independent of
energy in the wide-band limit approximation. Assuming weak coupling
between the sites and the reservoirs, we employ free-evolving proton
operators, $a_{\sigma}(t) = e^{- i \bar{E}_{\sigma}(t-t_1)}
a_{\sigma}(t_1),$ to calculate the anticommutator in Eq.
(\ref{nsigma2}). Here
\begin{equation}
\bar{E}_{\sigma} = E_{\sigma} - U _{q}(\phi+\phi_{\sigma}) - U_{\rm
con}(\phi+\phi_{\sigma}), \label{barEsigma}
\end{equation}
is the total eigenenergy of the proton on the site $\sigma$
including contributions of the stator charge potential, $U_q
(\phi),$ and the confining potential, $U_{\rm con}(\phi)$. As a
result, we derive a set of rate equations for the populations
$n_{\sigma}$, partially averaged over the Fermi distributions of the
proton reservoirs:
\begin{equation}
\dot{n}_{\sigma} + \sum_{\alpha} \Gamma_{\alpha\sigma}(\phi) \
n_{\sigma} \ =\  \sum_{\alpha} \Gamma_{\alpha\sigma}(\phi)\
f_{\alpha}(\bar{E}_{\sigma}). \label{rateEq}
\end{equation}
Hereafter we drop the averaging brackets $\langle ...\rangle$ and
introduce the angular-dependent transition coefficients,
$$\Gamma_{\alpha\sigma}(\phi)\,=\,\Gamma_{\alpha}\; w_{\alpha}^2(\phi
(t)+\phi_{\sigma}).$$ The rate equations (\ref{rateEq}) replace the
phenomenological Markovian equations
\cite{Dimroth99,Xing06,WangOster02}, which are usually employed for
a description of the loading and unloading of the rotor sites. The
characteristic time of proton transfer to and out of the
proton-binding sites \cite{Elston98,Aksim04} is much shorter than
the time scale of the rotation angle $\phi$. Accordingly, we can
average the stochastic Eq.~(\ref{Langevin}) over fluctuations of the
proton reservoirs without averaging over the fluctuations of the
mechanical heat bath. The partially averaged proton population
$n_{\sigma}$, involved in Eq.~(\ref{Langevin}), depends on the local
fluctuating value of the rotor angle $\phi(t)$. In the next section
we solve numerically the stochastic equation (\ref{Langevin})
together with the system of rate equations (\ref{rateEq}) and
investigate various regimes of the rotary nanomotor.

\section{Results}
We consider a cylinder-shaped motor with radius $r_0$ = 3 nm and
height $h$ = 6 nm, rotating in a medium with viscosity coefficient
$\eta$ = 1 Pa$\cdot$s, which is 1000 times higher than the viscosity
of water. For a protein environment, surrounding the stator charge
$q|e|$ and having a dielectric constant $\varepsilon$ = 3, the
energy of the Coulomb coupling $U_{q}(\phi)$ (\ref{Uq}) is
proportional to the factor $ q e^2/(4\pi \varepsilon_0 \varepsilon)$
= $q \ \times$ 480 meV$\cdot$nm. The Debye screening radius is $r_s$
= 1 nm. We also assume that both, the stator charge and the end of
the source channel, are located in the rotor plane on the $y$-axis
with $\phi_S = \pi$, and that the stator charge is offset by the
minimum distance $d$ = 0.9 nm from the circumference of the rotor
ring. The drain channel is shifted to the right by the angle
$\phi_0: \phi_D = \pi - \phi_0.$ For the steepness $\lambda_r$
involved in the function $w_{\alpha }(\phi)$, we choose the value:
$\lambda_r$ = (0.25 nm)$^{-1}$. The confinement potential
(\ref{Usigmac}) is characterized by the parameters $U_c$ = 1160 meV,
$\lambda_c$ = 20. We assume that all proton-binding sites $(\sigma =
1,\ldots,12)$ have the same eigenenergy, $E_{\sigma}=E_0$. The
proton transport through the system occurs if the $\phi$-dependent
energy levels of the proton-binding sites, $E(\phi) = E_0 - U_{q} (
\phi ) - U_{\rm con}( \phi),$ located near the source and drain
half-channels, $\phi=\phi_S,\phi_D,$ fit the transport window:
\begin{equation}
\frac{V}{2}\  > \ E(\phi_S)\  > \ E(\phi_D)\  > -\  \frac{V}{2}. \label{transpo}
\end{equation}
The energy $E_0$ = $-$ 90 meV gives the values $E(\phi_D)$ = $-$ 84
meV, $E(\phi_S)$ = 100 meV, which meet these conditions at the
transmembrane potential $V$ = 250 meV. It is assumed that the time
scales for transitions of protons between the reservoirs and the
rotor sites are about $\Gamma_S^{-1} = \Gamma_{D}^{-1}$= 0.5 $\mu$s.

\subsection{Rotations and site depopulations}

In Fig.~2a we plot a number of full rotations, $\phi(t)/2\pi$, as a
function of time (in ms), obtained from the numerical solution of
the Langevin Eq.~(\ref{Langevin}), \textit{coupled} to the rate
Eqs.~(\ref{rateEq}), at the source-drain voltage $V $ = 250 meV, the
stator charge $|e|$ ($q$=1), and at temperature $T$ = 300 K. We take
into account here the constant load torque from the F$_1$ motor,
${\cal T}_{\rm ext}$~=~$-$ 41 pN$\cdot$nm, which is enough to
produce three ATP molecules per rotation of the F$_1$ motor.

In the initial state all sites are occupied with protons, and,
because of this, are electrically neutral. The process of site
depopulation is illustrated in Fig. 2b, where we plot the
probability to have the empty $\sigma$-site, $\bar{n}_{\sigma} = 1 -
n_{\sigma}, $ as a function of time for $\sigma = 1,\ldots,12.$ The
maximum value of $\bar{n}_{\sigma}$ on this graph corresponds to the
unoccupied proton-binding site $\sigma$. The site ``5" is
depopulated in the first turn, because it is initially in the
closest position to the drain channel. The empty site is attracted
by the positive stator charge, and this force turns the rotor an
angle $\pi/6$ (see Fig.~2a). The site ``5" is occupied again when it
reaches the source channel. In Fig.~2c we plot the number $N_S$ of
protons, transferred from the source reservoir, as a function of
time. It is evident from Fig.~2c that loading the site ``5" is
accompanied by the transfer of a proton from the source channel. The
depopulation of the site ``4" begins at the same time as the
depopulation of the site ``5". This process repeats over and over,
resulting in a continuous unidirectional rotation of the rotary
ring. The loaded motor makes a complete turn in a time interval
$\sim$ 0.25 ms, which corresponds to the frequency of rotations near
4 kHz; in so doing the system translocates about 12 protons down the
potential gradient of 250 meV.

It should be noted that here we have an asymmetric configuration,
where the energy level of the proton near the source channel is
higher than the proton energy near the drain outlet, $E(\phi_S) >
E(\phi_D) $. The energy drop,
\begin{equation}
\Delta E \;=\; E(\phi_S) - E(\phi_D)\; =\; U_q(\phi_D) -
U_q(\phi_S), \label{EnDrop}
\end{equation}
is \textit{directly} converted into mechanical energy by the F$_0$
motor. We note that $U_q(\phi) < 0 $ and $ U_{\rm con}(\phi_S) =
U_{\rm con}(\phi_D)$. At the opposite sign of the transmembrane
potential difference, $V < 0,$ the motor does not produce any
unidirectional rotation, since in this case the proton energies do
not fit the transport window (\ref{transpo}).

\subsection{Torque generation and shuttling}

In Fig.~3 the average rotational frequency, $\langle
\Omega\rangle/2\pi = \langle \dot{\phi}\rangle/2\pi$, shown in (a),
and the proton current, $I_S = N_S/\tau_R$, (b), averaged over the
time interval $\tau_R = 2 $ ms as well as over several realizations,
are presented as functions of the eigenenergy of the proton-binding
states on the rotor, $E_0$, at three values of the temperature, $T$
= 4.2 K (liquid $^{4}$He); $T$ = 77.2 K (liquid $^{14}$N); and $T$ =
300 K.  The other parameters, namely, the proton voltage, $V$ = 250
mV, the normalized stator charge, $q$ = 1, and the external torque,
${\cal T}_{\rm ext}$ = $-$ 41 pN$\cdot$nm, correspond to the loaded
F$_0$ motor. At resonant values of $E_0: E_0 \sim$ $-$ 90 meV, and
at high enough temperatures, $T >$ 50 K, the motor works in a
\textit{torque-generating} regime characterized by a positive (CW)
direction of rotation, $\langle \Omega\rangle > 0$, and positive
current, $I_S
> 0$, when protons flow downhill,  from the source to the drain
reservoir. In this regime \textit{both} mechanisms, the power stroke
and the Brownian ratchet, contribute to the torque generation.

At low temperatures, e.g. when $T$ = 4.2 K, the ratchet mechanism
practically turns off, and the system is not able to produce enough
torque to execute the full-circle rotation in the presence of the
counteracting load torque ${\cal T}_{\rm ext}$. In this
\textit{shuttling} mode, one proton-binding site oscillates back and
forth between the source and drain channels, translocating the
protons ($I_S > 0$), but no unidirectional rotation is generated
$\langle \Omega \rangle$ = 0).

When $E_0 <$ $-$ 160 meV the system is out of the transport windows,
and the proton-binding sites on the rotor are always populated.
Then, the rotor follows the negative external torque and rotates in
the CCW direction without transferring protons (i.e.,
$\langle\Omega\rangle < 0,\ I_S \simeq 0$).

\subsection{Torque generation and proton pumping}

In Fig.~4 we show the dependence of  (a) the average speed of
rotations, $\langle \Omega\rangle/2\pi$, and (b) the average
particle current, $I_S$ (average number of protons transferred from
the source channel per one millisecond), on the voltage $V$ for the
whole range of the CCW-directed external torque ${\cal T}_{\rm
ext}$, from zero to $-$ 120 pN$\cdot$nm, at $E_0 =$ $-$ 90 meV, and
$T$ = 300 K. For values of the external torque between zero and $-$
80 pN$\cdot$nm and at a sufficiently high voltage, $V > 220$ meV,
the motor performs a CW rotation (Fig.~4a) with a maximum frequency
$\sim$10~kHz (the \textit{torque-generating} mode). In the process,
protons flow downhill, from the source $(\mu_S = V/2)$ to the drain
channel $(\mu_D = -V/2)$. The CCW- directed external torque hampers
this motion, and when $|{\cal T}_{\rm ext}| >$ 80 pN$\cdot$nm the
rotation stops (Fig.~4a).

In the regime of ATP hydrolysis \cite{Dimroth06}, the F$_1$ motor
produces a sufficient torque to drive the rotation of the F$_0$
motor in the reverse (CCW) direction. In our case this regime takes
place when $|{\cal T}_{\rm ext}| >$ 50 pN$\cdot$nm, provided that
the transmembrane potential is small enough, $V < 170$ meV
(Fig.~4a). It is evident from Fig.~4b that in this range of
parameters the system works as a proton pump ($I_S < 0$), which
translocates protons against the gradient of the electrochemical
potential with a pumping efficiency
\begin{equation}
{\rm Eff}_{\rm pump} \;=\; \frac{I_S\cdot V}{{\cal T}_{\rm ext}
\cdot \langle \Omega \rangle } \;\sim \; 20\% \label{EffPump}
\end{equation}
when $V$ = 100 meV and ${\cal T}_{\rm ext}$ = $-$ 108 pN$\cdot$nm.

Note, that the voltage threshold for the torque-generating regime
($V
> $ 210 meV when ${\cal T}_{\rm ext}$ = $-$ 41 pN$\cdot$nm)
corresponding to the value of the electrochemical potential
difference across the bacterial membrane and the inner mitochondrial
membrane, is \textit{not} preassigned a priori, but it is naturally
determined by the configuration parameters of the model; for
example, by the charge and the position of the stator as well as by
the value of the load torque necessary to drive the ATP synthesis.

\subsection{Contributions from the Brownian ratchet and power stroke
components}

The model studied here can be considered as a combination of
Brownian ratchet \textit{and} power stroke components
\cite{OsterWang03,WangOster02}. The Brownian ratchet harnesses the
energy of thermal fluctuations, biased by chemical reactions, to
generate a unidirectional rotation. In our case, the ratchet
component is introduced via a fluctuation force $\xi$, working in
combination with the processes of loading and unloading protons in
the presence of the confining potential $U_{\rm con}$ (see
Eqs.~(\ref{Langevin}),\,(\ref{rateEq})). It is expected that the
ratchet contribution should be significantly diminished at low
temperatures.

The power stroke component is due to the Coulomb attraction between the positive stator charge, $q|e|$, and the negatively charged
unoccupied rotor site. This part of the torque-generating process is proportional to the coefficient $q$, which is the stator
charge, measured in units of $|e|$. For the torque-generating and pumping process in the F$_0$ motor of \textit{E.~coli}, the
importance of the residue $c$Arg-210, carrying a positive charge,  has been emphasized in Refs. \cite{Elston98,Cain89}.

In Fig.~5, taking the values $d$ = 0.9 nm, $E_0$ = $-$ 90~meV,  and
V = 250 meV, we plot: (a) the average torque $\langle {\cal
T}\rangle$ generated by the isolated motor (at $|{\cal T}_{\rm
ext}|$ = 0), and (b) the efficiency \cite{Berg03},
\begin{equation}
{\rm Eff} \;= \;\frac{\langle {\cal T}\rangle\cdot\langle \Omega
\rangle}{I_S \cdot V}, \end{equation} \label{Eff}
 of the system as a function of the stator charge
$q$, normalized by $|e|$, at low ($T$~=~4.2 K, dashed blue line), intermediate ($T$ = 77.2 K, dash-dotted green line), and high
($T$ = 300 K, continuous red line) temperatures.

In the pure \textit{Brownian ratchet regime} (no stator charge, $q$~=~0) the system generates a pronounced torque, $\langle {\cal
T}\rangle \sim$ 12 pN$\cdot$nm, if the temperature is high, $T$ = 300 K. However, this torque is not enough to overcome the load
torque, ${\cal T}_{\rm ext}$ = $-$ 41 pN$\cdot$nm, which is necessary for ATP synthesis. For instance, at liquid Helium
temperatures ($T$ = 4.2 K) the Brownian ratchet component is strongly suppressed.

Nevertheless, the \textit{power stroke mechanism} can generate the torque $\langle{\cal T}\rangle_{\rm max} \simeq$ 60
pN$\cdot$nm, which is higher than the load torque from the F$_1$ motor. The isolated F$_0$ motor demonstrates this peak value of
the torque and the efficiency, $ {\rm Eff}_{\rm max} \simeq 80 \%,$ at the stator charge near $|e|$ ($q \simeq 1$).

It follows from Fig.~3 that the power stroke component alone is not able to drive ATP synthesis, because at the load conditions,
${\cal T}_{\rm ext}$ = $-$ 41 pN$\cdot$nm, and at $T$ = 4.2 K, the system operates in the \textit{shuttling mode}, with a proton
current but generating no unidirectional motion.
 Notice also that despite the presence of
a medium with high viscosity coefficient, $\eta$ = 1 Pa$\cdot$s =
1000$\,\cdot\, \eta_{\rm water}$, and despite the very small size,
$r_0$ = 3 nm, the motor performs quite well and generates a
significant torque at physiologically reasonable parameters.

Note that, because of the asymmetric configuration, this motor does not produce a unidirectional torque for a {\it negative}
electrochemical potential difference $(V < 0)$, but it switches the direction of rotation if the drain channel (see Fig.~1) is
placed on the other side of the source channel (at $\phi_D = \pi + \phi_0$). This feature, as well as the nanoelectromechanical
method as a whole, can be useful for explaining the {\it switching} ability of bacterial flagellar motors, which allows the
chemotaxis of bacteria \cite{Berg03,Chemotaxis}.

\section{Conclusions}

We have examined a simple nanoelectromechanical model of a proton-driven rotary nanomotor which mimics the basic operating
principles and the configuration of the F$_0$-motor of ATP synthase in bacterial and mitochondrial membranes. Treating ATP
synthase as a nonequilibrium open system, coupled to two proton reservoirs, we have derived a set of rate equations, which
describes the loading and unloading of the proton-binding sites on the rotor portion of the motor.

At normal conditions, the isolated motor generates a torque of about 60 pN$\cdot$nm with efficiency near 80\%, whereas the motor
working against a constant load, created by the F$_1$ portion, exhibits a unidirectional rotation (Fig.~2) with frequency $\sim$ 4
kHz. It is shown that, depending on the temperature $T$, the external torque ${\cal T}_{\rm ext}$, the proton voltage build-up V,
and the energy $E_0$ of the rotor sites, the system operates in \textit{three} different regimes: the \textit{torque-generating}
mode; the \textit{shuttling} regime; and the \textit{pumping} mode (see Fig.~3 and Fig.~4).

In the \textit{torque-generating} mode, energy  (stored in the
gradient of the proton electrochemical potential) is converted into
mechanical energy of the rotor. This energy transfer drives the
unidirectional rotation of the motor, $\langle\Omega\rangle > 0,$
and the positive particle current, $I_S > 0$, corresponding to the
downhill flow of protons from the periplasmic (P) to the cytoplasmic
(N) side of the membrane.

In the \textit{shuttling} regime, which takes place at low temperatures, the rotor vibrates with a small amplitude near the
initial point without performing a full circle. In doing so,  the proton-binding site, located between the source and drain
half-channels, works as a nanomechanical shuttle carrying protons from the P-side to the N-side of the membrane. This regime is
distinguished by zero-frequency rotations, $\langle\Omega\rangle = 0,$ and by a positive proton current, $I_S > 0$.

In the \textit{pumping} mode \cite{Elston98,Dimroth06}, ATP synthase
operates in reverse,  where ATP hydrolysis drives rotations of the
F$_1$ motor. This external torque, ${\cal T}_{\rm ext}$, is
transmitted to the F$_0$ motor, which pumps protons uphill, from the
N to P membrane side. In our case, the CCW-directed external torque,
${\cal T}_{\rm ext}$ = $-$ 120 pN$\cdot$nm, is enough to
translocate, in one millisecond, about 90 protons against the
electrochemical gradient of 100 meV (see Fig.~4b).

We have studied the performance of the model at different
temperatures and found that this performance depends significantly
on the Coulomb interaction, $U_q$, between the positive stator
charge and the negatively charged empty rotor site. The torque
produced by the Brownian ratchet mechanism alone is not sufficient
to overcome the load torque from the F$_1$ motor. A key feature of
the model is that, because of Coulomb coupling to the stator charge,
the energy of the proton on the rotor site, which is close to the
source half-channel, $E(\phi_S)$, is \textit{higher}, than the
proton energy of the site located near the outlet of the drain
channel, $E(\phi_D)$.

For positive proton voltage $V$, when the electrochemical potential of the source, $\mu_S = V/2,$ exceeds the potential of the
drain, $\mu_D = -V/2,$ the rotor sites can be loaded with protons at the source and unloaded at the drain, if the energies
$E(\phi_S)$ and $E(\phi_D)$ fit the transport window: $\mu_S > E(\phi_S) > E(\phi_D)
> \mu_D.$ This energy difference, $\Delta E = E(\phi_S)-E(\phi_S) =
U_q(\phi_D) - U_q(\phi_S), $ is \textit{directly} converted into mechanical energy. It is shown, however, that at low
temperatures, when the Brownian ratchet component is suppressed, the motor works in the shuttling mode, and no torque is
generated. This means that, in agreement with previous results \cite{Elston98,Aksim04}, the power stroke and the Brownian ratchet
components should work together for the efficient operation of the system. These conclusions can be applied both to biological
rotary motors and to artificial nanoengines. Notice also that an efficient and powerful future synthetic motor based on this
bio-inspired design would be expected to withstand a more severe environment,  in particular, much lower temperatures, than its
biological counterparts. The present design can be compared with modern industrial electrical motors having an efficiency about
90\%. However, it should be emphasized that this efficiency decreases drastically when decreasing the size of the motors.
Nano-scale motors operating in a warm and very viscous medium, even with an efficiency 80\%, are worth building and studying as a
function of various operating conditions.

\section{Acknowledgement}

This work was supported in part by  the National Security Agency
(NSA), Laboratory of Physical Science (LPS), Army Research Office
(ARO), National Science Foundation (NSF) grant No. EIA-0130383,
JSPS-RFBR 06-02-91200, and Core-to-Core (CTC)
 program supported by the Japan Society for Promotion of Science (JSPS).
 S.S. acknowledges support from the EPSRC ARF No. EP/D072581/1 and AQDJJ
network-programme. L.M. is partially supported by the NSF NIRT,
grant ECS-0609146.

\newpage

\begin{figure}[ht]
\includegraphics[width=11.0cm, height=14.0cm]{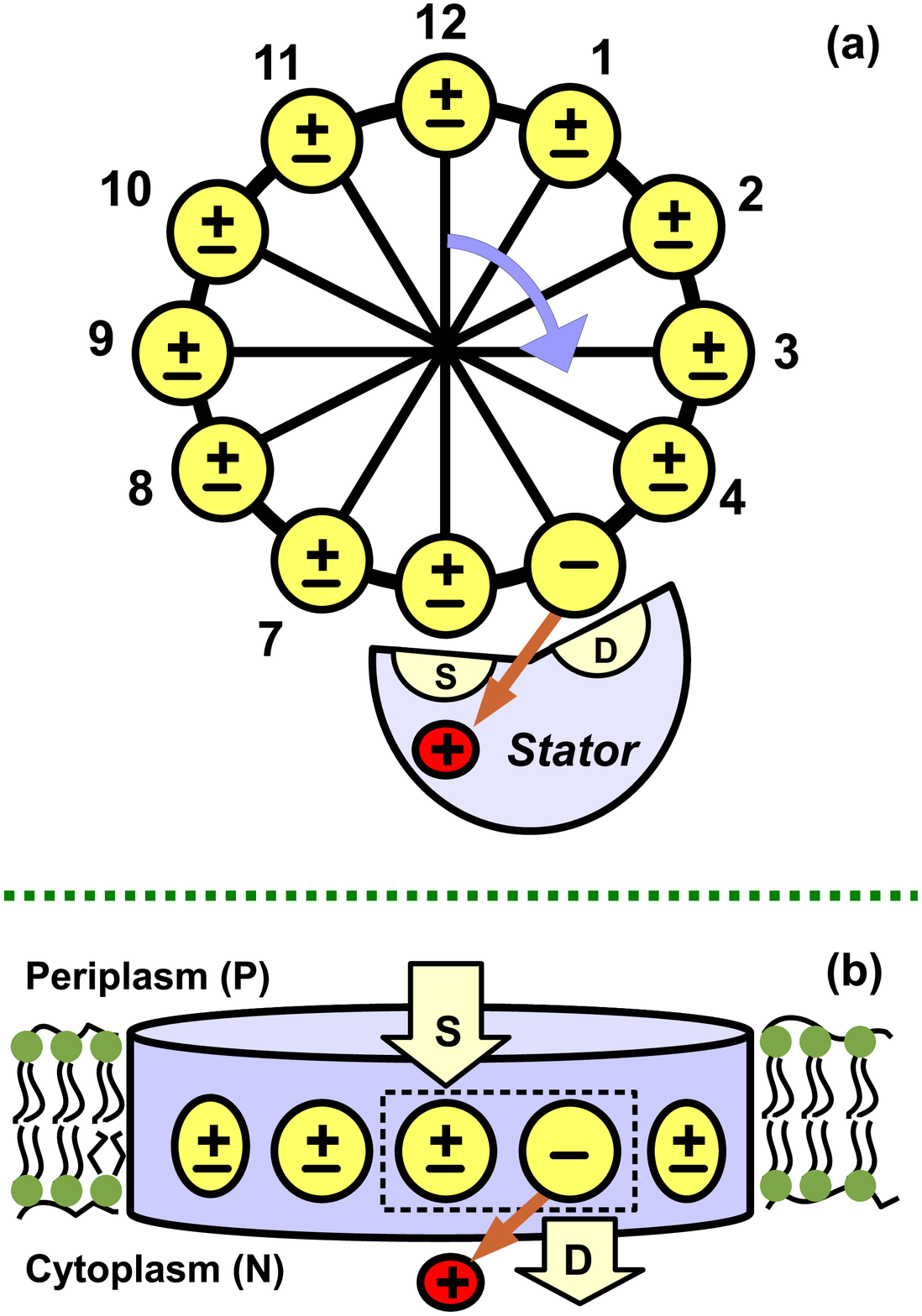}
\vspace*{0cm} \caption{(Color online) Schematic diagram of a rotary
nanomotor powered by an electrochemical potential gradient. (a)
shows the top view from the periplasmic (P) side of the membrane;
(b) shows the side view. Panel (a) shows the rotor, which has twelve
sites which rotate as  a Ferris wheel. In the example shown in (a),
the site number five provides a proton to the drain ($D$) channel of
the stator, and the proton then moves to the cytoplasm, as shown in
(b). This now negatively-charged site number five in (a) feels a
Coulomb attraction from the positive (red) stator charge, which
propels the wheel to rotate an angle $\phi_0 = 30^{\circ}.$ After
site number five moves to the location of the next site, the source
($S$) donates a proton to that site, neutralizing it, and the
30$^{\circ}$ rotation process is ready to start all over again. In
(b), the protons, indicated by ``+", move from the periplasm ($P$),
which acts as a source ($S$), to the cytoplasm ($N$), acting as a
proton drain ($D$), on the other side of the membrane. In (a) and
(b), an orange arrow represents the Coulomb force between the empty
rotor site and the positive stator charge.}
\end{figure}

\begin{figure}
\includegraphics[width=26.0cm, height=13.0cm]{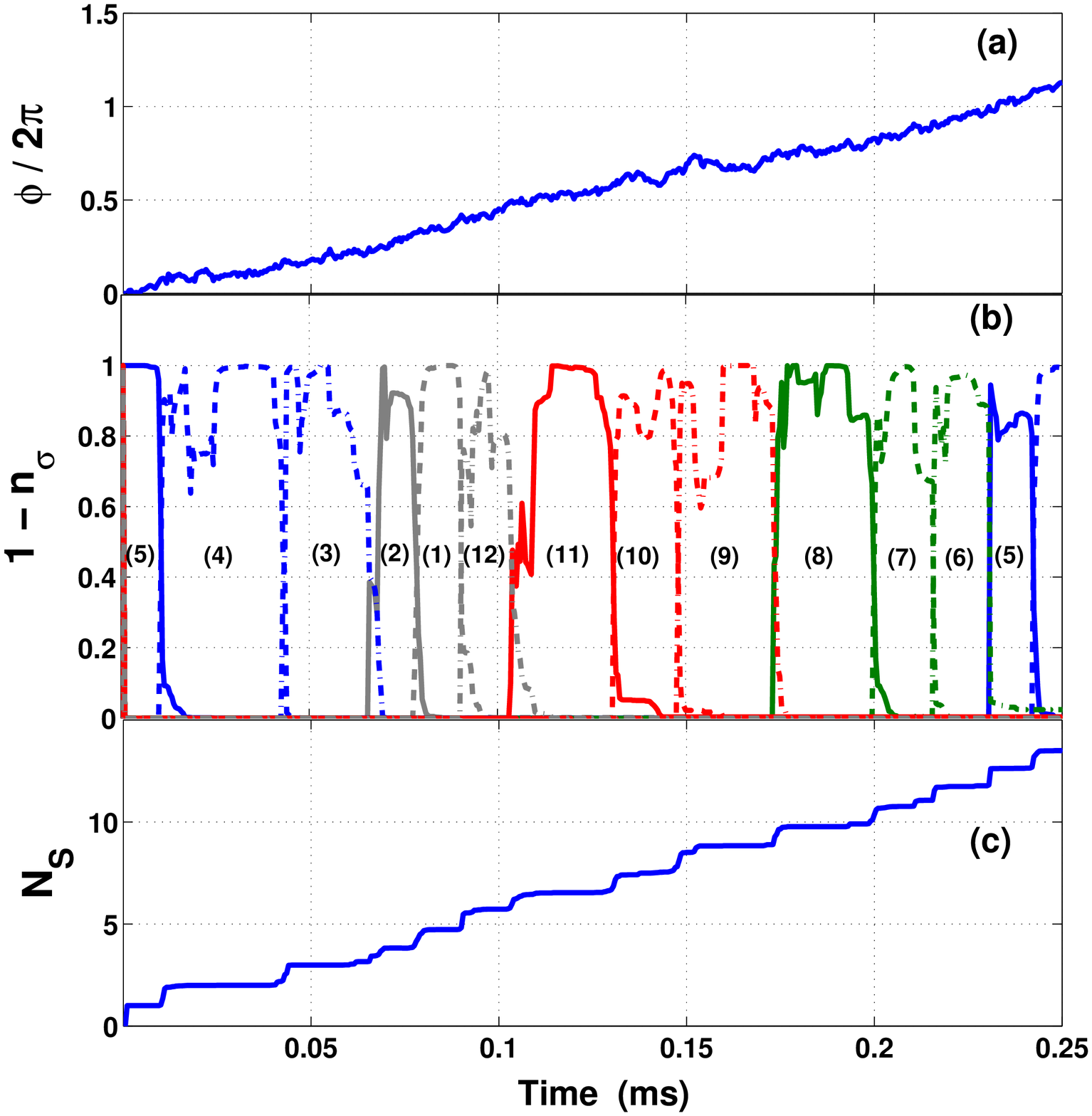}
\vspace*{0cm} \caption{(Color online) (a) Time dependence of the number of full rotations $\phi (t) /2\pi$ at $V$ = 250 meV,
${\cal T}_{\rm ext}$ = $-$ 41 pN$\cdot$nm, $E_0$ = $-$ 90 meV, and at $T$ = 300 K; (b) Depopulations of the proton-binding sites,
$\bar{n}_{\sigma} = (1 - n_{\sigma}),\ \sigma = 1,\ldots,12,$ versus time (in ms); (c) The number of protons, $N_S$, transferred
from the source to the drain reservoir, versus time (in ms). Notice the periodicity in (b). It follows from Fig.~2a, that the
motor performs a bit more than one full CW rotation for a period of time $\sim$ 0.25 ms. This rotation is accompanied by the
cyclic loading and unloading of the proton-binding sites on the rotor (Fig.~2b), starting with the site number five (see Figs.~1a
and 2b). All sites were populated at the initial time $t=0$. The rotor works as a Ferris wheel carrying protons from the source
reservoir $S$ to the drain $D$. The number of protons, $N_S$, translocated form the source to the drain, increases stepwise during
the rotation of the wheel (Fig.~2c). Almost 12 protons are transferred through the 12-site system after the full circle, as seen
in Fig.~2c.}
\end{figure}

\begin{figure}
\includegraphics[width=25.0cm,  height=13.0cm]{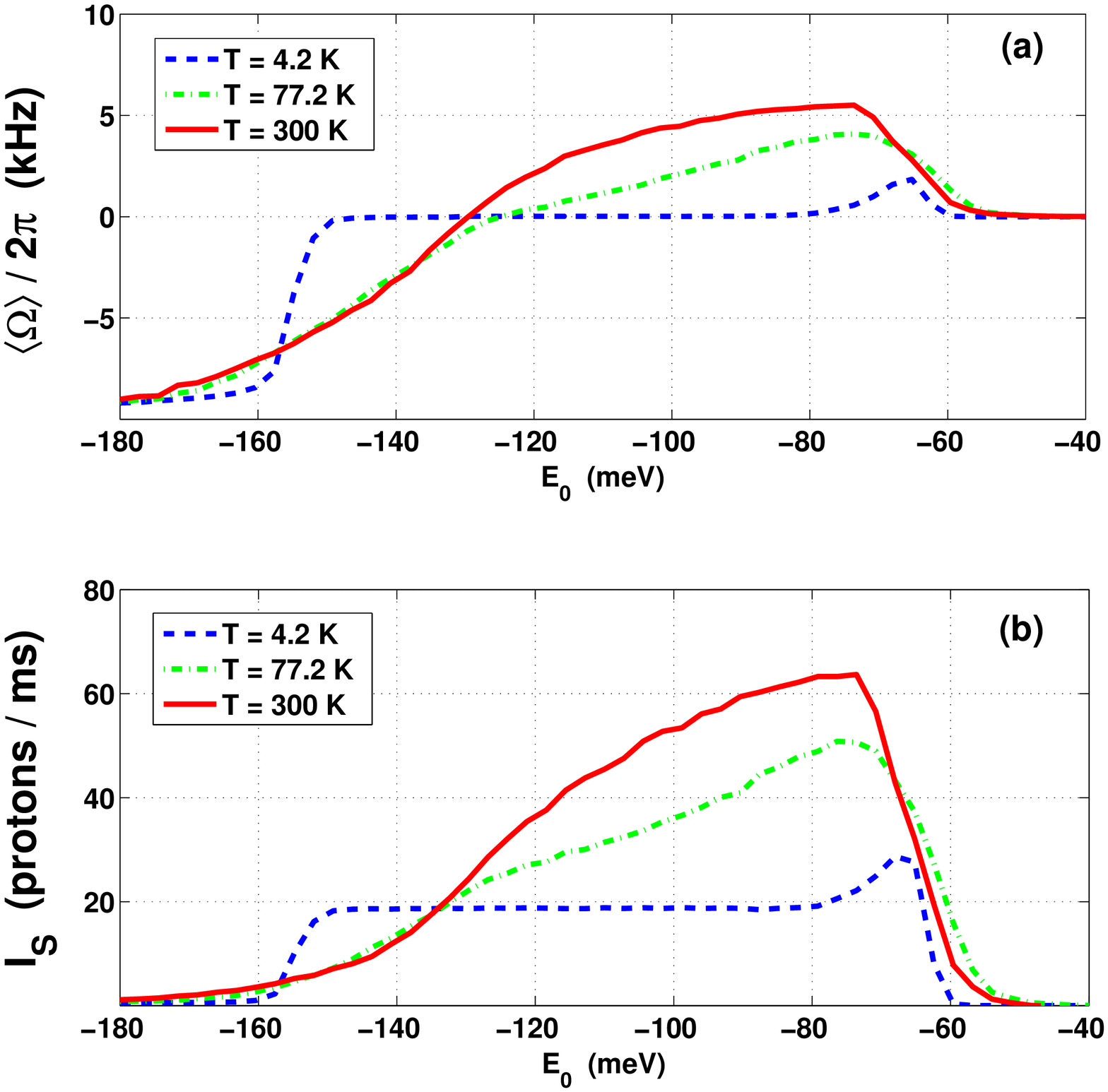}
\vspace*{0cm} \caption{ (Color online) (a) The average frequency of rotations, $\langle \Omega\rangle/2\pi$ (in kHz), and (b) the
average proton current $I_S$ (the number of protons, transferred through the system in one millisecond) as functions of the energy
of the rotor sites, $E_0$ (in meV), at $V$ = 250 meV, ${\cal T}_{\rm ext}$~=~$-$~41~pN$\cdot$nm, and at different temperatures:
$T$ = 4.2 K; $T$ = 77.2 K; and $T$ = 300 K. In the \textit{torque-generating} regime ($T$ = 300  K, $E_0 \sim - 80$ meV), despite
the counteracting load torque ${\cal T}_{\rm ext}$, the system rotates CW, $\langle \Omega \rangle > 0,$ performing about 5 full
circles (Fig.~3a, red continuous line) and carrying more than 60 protons from the source to the drain in one millisecond
(Fig.~3b). However, at very low temperatures, $T$ = 4.2 K, the Brownian ratchet component of the torque-generating process turns
off, and the rotor is not able to overcome the external torque. The system is now in the \textit{shuttling} regime oscillating
near a fixed position with zero average frequency of rotations, $\langle \Omega \rangle = 0$, but carrying about 20 protons per
millisecond (Figs.~3a and~3b, flat plateaus in the blue dashed curves). }
\end{figure}

\newpage

\begin{figure}
\includegraphics[width=25.0cm,  height=13.0cm]{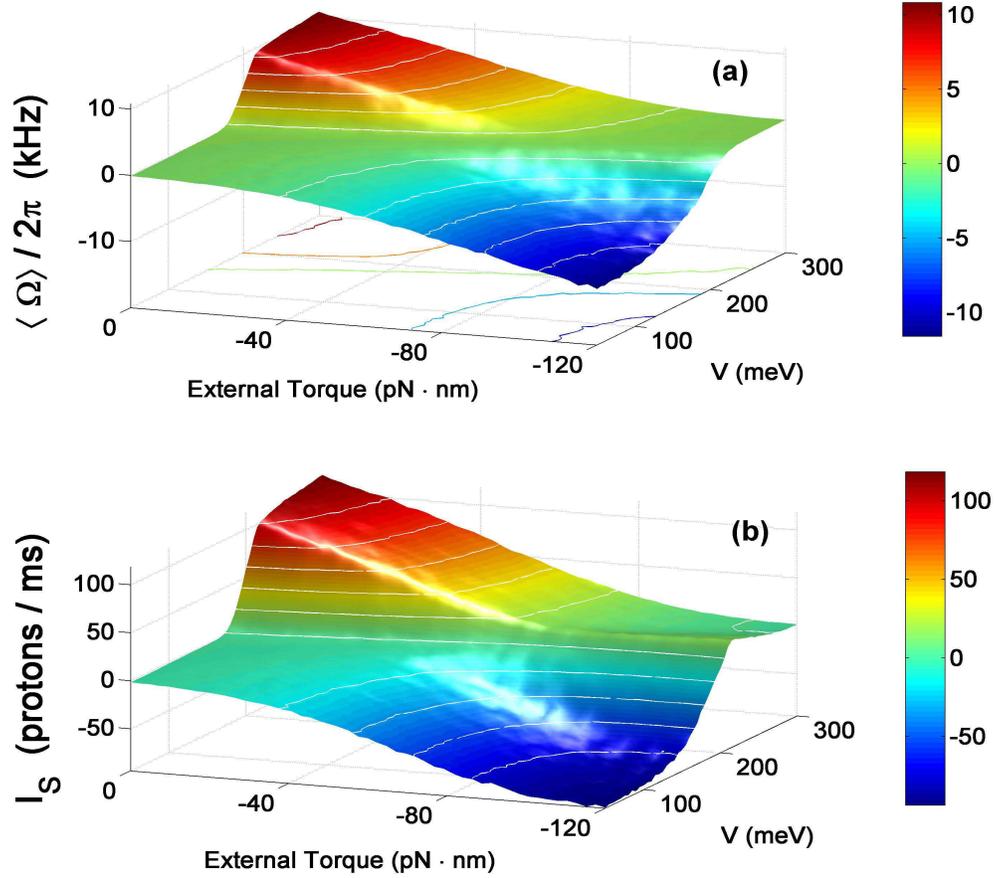}
\vspace*{0cm} \caption{(Color online) (a) the average speed of rotations $\langle \Omega \rangle/2\pi$ (in kHz), and (b) the
average proton current, $I_S$, as functions of the proton voltage, $V$ (in meV), and the external torque, ${\cal T}_{\rm ext}$ (in
pN$\cdot$nm), at $E_0$ = $-$ 90 meV, and $T$ = 300 K. Figure 4a shows that for a low enough counteracting load torque, $|{\cal
T}_{\rm ext}| < 80$ pN$\cdot$nm, and at a high proton voltage, $V > 220$ meV, the motor rotates clockwise with a maximum frequency
of about 10 kHz. This rotation is accompanied by the positive proton current reaching the value $I_S \sim 100$ protons/ms
(Fig.~4b). The blue (negative) regions in Fig.~4 correspond to the regime of ATP hydrolysis (\textit{pumping} mode), when at low
voltages, $V~<~170$ meV, the external torque, $|{\cal T}_{\rm ext}|~>~50$ pN$\cdot$nm, is powerful enough to drive the motor
rotation in the CCW direction ($\langle \Omega \rangle~<~0$), pumping protons ($I_S~<~0$)  from the drain to the source against
the proton electrochemical gradient $V$ (Fig.~4b). }
\end{figure}

\begin{figure}
\includegraphics[width=25.0cm,  height=12.0cm]{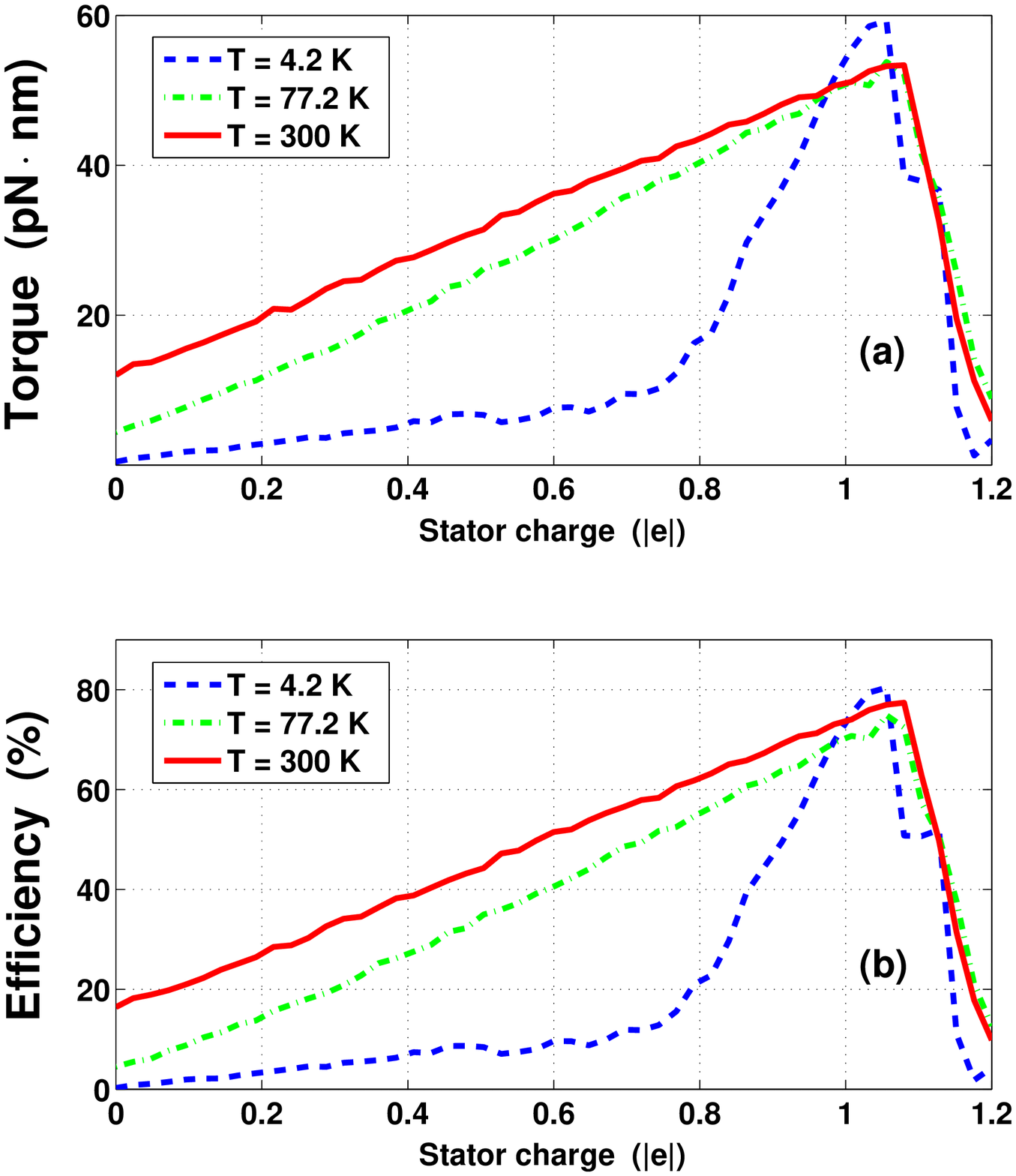}
\vspace*{0cm} \caption{(Color online) (a) Average torque, $\langle {\cal T}\rangle$ (in pN$\cdot$nm), and efficiency, Eff, of the
isolated motor (${\cal T}_{\rm ext}$ = 0) as functions of the normalized stator charge $q$ (in units of $|e|$) for $V$ = 250 meV,
$E_0$ = $-$ 90 meV, and for three different temperatures: $T$~=~4.2 K (blue dashed line); $T$~=~77.2 K (green dash-dotted line);
and $T$~=~300 K (red continuous line). Without the stator charge (Fig.~5a, red continuous line, $T$ = 300 K, $q \,\sim\, 0$) the
ratchet mechanism produces the torque $\sim$ 12 pN$\cdot$nm, which is not enough to drive ATP synthesis. Near the optimal value of
the stator charge ($q\,\sim\, 1$), but at very low temperatures, $T$ = 4.2 K, the isolated motor generates the torque 60
pN$\cdot$nm by means of the power stroke mechanism (Fig.~5a, blue dashed line). This torque is higher than the CCW torque ($ -\,
41$ pN$\cdot$nm) necessary for ATP production. However, at these conditions, the F$_0$ motor coupled to the load (${\cal T}_{\rm
ext} = -\, 41$ pN$\cdot$nm) switches from the torque-generating mode to the shuttling regime, as shown in Fig.~3, and no average
torque is produced. Figure 5b demonstrates that the efficiency of the motor can reach $\simeq 80 \%,$ when \textit{both}
components contribute to the torque-generating process. }
\end{figure}

\end{document}